\documentclass[a4paper,11pt]{article}
\usepackage{jcappub} % for details on the use of the package, please see the JINST-author-manual
\usepackage{lineno}
\usepackage{xcolor}
\usepackage{float}

\usepackage{makecell}
\usepackage{multirow}
\usepackage{tablefootnote}
\usepackage{array}
\usepackage{silence}
\arxivnumber{2412.09064} % Only if you have one
\title{A comprehensive numerical study on four categories of holographic dark energy models}

\author[a]{Jun-Xian Li,}
\author[a]{Shuang Wang, \note{Corresponding author.}$^{1}$}
\affiliation[a]{School of Physics and Astronomy, Sun Yat-Sen University, Zhuhai, China}

\emailAdd{lijx389@mail2.sysu.edu.cn}
\emailAdd{wangshuang@mail.sysu.edu.cn}

\abstract{Holographic dark energy (HDE), which arises from a theoretical attempt to apply the holographic principle (HP) to the dark energy (DE) problem, has attracted significant attention over the past two decades. We perform a comprehensive numerical study on HDE models that can be classified into four categories: 1) HDE models with other characteristic length scale, 2) HDE models with extended Hubble scale, 3) HDE models with dark sector interaction, 4) HDE models with modified black hole entropy. For theoretical models, we select seven representative models, including the original HDE (OHDE) model, Ricci HDE (RDE) model, generalized Ricci HDE (GRDE) model, interacting HDE (IHDE1 and IHDE2) models, Tsallis HDE (THDE) model, and Barrow HDE (BHDE) model. For cosmological data, we use the Baryon Acoustic Oscillation (BAO) data from the Dark Energy Spectroscopic Instrument (DESI) 2024 measurements, the Cosmic Microwave Background (CMB) distance priors data from the Planck 2018, and the type Ia supernovae (SNe) data from the PantheonPlus compilation. Using $\chi^2$ statistic and Bayesian evidence, we compare these HDE models with current observational data. It is found that: 1) The $\Lambda$CDM remains the most competitive model, while the RDE model is ruled out. 2) HDE models with dark sector interaction perform the worst across the four categories, indicating that the interaction term is not favored under the framework of HDE. 3) The other three categories show comparable performance. The OHDE model performs better in the BAO+CMB dataset, and the HDE models with modified black hole entropy perform better in the BAO+CMB+SN dataset. 4) HDE models with the future event horizon exhibit significant discrepancies in parameter space across datasets. The BAO+CMB dataset favors a phantom-like HDE, whereas the BAO+CMB+SN leads to an equation of state (EoS) much closer to the cosmological constant.}

\begin{document}
\maketitle
\flushbottom

\section{Introduction}
\label{sec:intro}

In 1998, observations of Type Ia supernovae led to the groundbreaking discovery of cosmic acceleration, revealing the existence of a mysterious dominant component: dark energy (DE) \cite{SupernovaSearchTeam:1998fmf, SupernovaCosmologyProject:1998vns}. The $\Lambda$ Cold Dark Matter ($\Lambda$CDM) model, which interprets DE as a cosmological constant, has been widely regarded as the standard model of modern cosmology. However, the $\Lambda$CDM model still faces some theoretical challenges, such as the fine-tuning problem and the coincidence problem \cite{Weinberg:1988cp, Carroll:2000fy, Peebles:2002gy, Padmanabhan:2002ji, Copeland:2006wr, Frieman:2008sn, Silvestri:2009hh, Li:2011sd, Bamba:2012cp, Li:2012dt, Bull:2015stt, Bullock:2017xww, Filippi:2020kii, SolaPeracaula:2022hpd}. Moreover, in terms of observation, the $\Lambda$CDM model faces the "Hubble tension" problem, a discrepancy between the directly measured current cosmic expansion rate and its inferred value from early-universe observations \cite{Verde:2019ivm, Perivolaropoulos:2021jda, Schoneberg:2021qvd, DiValentino:2021izs, Abdalla:2022yfr, Kamionkowski:2022pkx, Poulin:2023lkg, Rong-Gen:2023dcz, Efstathiou:2024dvn}.

Recently, the $\Lambda$CDM model has encountered a new observational challenge. The DESI collaboration released its first-year data on BAO \cite{DESI:2024mwx}. By analyzing the observational data with the Chevallier-Polarski-Linder (CPL) parameterization \cite{Chevallier:2000qy, Linder:2002et}, the DESI collaboration found the sign of time-varying DE EoS. Initially, this sign was observed with a significance of 2.6$\sigma$ in a combined analysis of DESI BAO and CMB data. After including various SNe Ia datasets, the discrepancy intensified, with a significance ranging from 2.5$\sigma$ to 3.9$\sigma$. Because of these theoretical and observational challenges to the $\Lambda$CDM model, it is necessary to explore dynamical DE models.

Dynamical DE models suggest that the universe's accelerated expansion is driven by a time-evolving DE, characterized by an EoS $w(z)$ that varies with redshift $z$. Notable dynamical DE models had been proposed, such as quintessence \cite{Wetterich:1987fm, Ratra:1987rm},  phantom \cite{Caldwell:1999ew}, k-essence \cite{Chiba:1999ka, Armendariz-Picon:2000nqq, Armendariz-Picon:2000ulo}, Chaplygin gas \cite{Kamenshchik:2001cp, Bilic:2001cg, Bento:2002ps}, etc. For a comprehensive overview of recent developments in this area, we refer the reader to Refs. \cite{Yoo:2012ug, Tsujikawa:2013fta, Joyce:2016vqv, Arun:2017uaw, Bahamonde:2017ize, Kase:2018aps, Avsajanishvili:2023jcl, Gialamas:2024lyw, Li:2024qso, Dhawan:2024gqy}.

As a prominent example of dynamical DE, HDE originates from a theoretical attempt to apply the holographic principle (HP) \cite{tHooft:1993dmi, Susskind:1994vu} to the DE problem. The HP states that all the information contained within a volume of space can be encoded on the boundary of that space, much like a hologram. It means that the energy density of DE $\rho_{de}$ can also be described by quantities on the boundary of the universe, including the characteristic length scale $L$ and the reduced Planck mass $M_P^2 \equiv 1/(8\pi G)$. 
Based on the dimensional analysis, we have \cite{Wang:2016och}:
\begin{equation}
\label{rhode-exp}
    \rho_{de} = C_1 M_P^4 + C_2 M_P^2 L^{-2} + C_3 L^{-4} + \cdots,
\end{equation}
where $C_1$, $C_2$, $C_3$ are constant parameters. However, the leading term is $10^{120}$ times larger than the cosmological observations \cite{Peebles:2002gy}, so this term should be excluded. Moreover, compared with the second term, the third term and the other terms are negligible and can be disregarded. Therefore, the expression of $\rho_{de}$ can be written as \cite{Wang:2016och}:
\begin{equation}
\label{rhode}
    \rho_{de} = 3C^2 M_p^2 L^{-2},
\end{equation}
where $C$ is a dimensionless constant parameter. It is important to emphasize that Eq. (\ref{rhode}) serves as the foundational expression for the HDE energy density.

Theoretically, HDE exhibits a strong dependence on the choice of characteristic length scale $L$. Although the Hubble scale $L = 1/H$ is a natural choice for the characteristic length scale, it has been demonstrated that this choice leads to an incorrect EoS for DE \cite{Horava:2000tb, Thomas:2002pq, Hsu:2004ri}. Over the past two decades, extensive studies on HDE had led to various theoretical models. These models can be divided into four main categories \cite{Wang:2023gov}: (1)HDE model with other characteristic length scale \cite{Li:2004rb, Guberina:2005mp, Cai:2007us,  Wei:2007ty, Huang:2012xma}; (2)HDE models with extended Hubble scale \cite{Nojiri:2005pu, Gao:2007ep, Granda:2008dk, Gong:2009dc}; (3)HDE models with dark sector interaction \cite{Zimdahl:2007zz, Wang:2005jx, Setare:2006wh, Xu:2009ys, Wang:2016lxa}; (4)HDE models with modified black hole entropy \cite{Tavayef:2018xwx, Saridakis:2018unr, DAgostino:2019wko, Saridakis:2020zol, Srivastava:2020cyk, Adhikary:2021xym, Drepanou:2021jiv}.

On the other hand, the numerical studies of various HDE models had also attracted considerable attention \cite{Huang:2004wt, Kao:2005xp, Feng:2007wn, Zhang:2007zga, Lu:2009iv, Zhang:2009un, Li:2009bn, Duran:2010hi, Wang:2010kwa, Li:2010xjz, Fu:2011ab, Zhang:2015rha, Cui:2015oda, Xu:2016grp, Wen:2017aaa, Basilakos:2017rgc, Li:2019bqb, daSilva:2020bdc, Dai:2020rfo, Dabrowski:2020atl, Anagnostopoulos:2020ctz, Hernandez-Almada:2021aiw, Qiu:2021cww, Rezaei:2022bkb, Oliveros:2022biu, Luciano:2023wtx, Feng:2023cbl, Denkiewicz:2023hyj, Nakarachinda:2023jko, Fang:2024yni, Adolf:2024twn}.
It should be emphasized that, previous studies mainly focused on the individual HDE models.
For examples, Ref. \cite{Tang:2024gtq} investigated the cosmological constraints on the OHDE model, and found that it performs a litter better than the $\Lambda$CDM model for the combined datasets including Planck CMB angular and weak lensing power spectra, Atacama Cosmology Telescope temperature power spectra, BAO, redshift-space distortion (RSD) and Cepheids-Supernovae measurement from SH0ES team (R22). However, after adding the PantheonPlus SN data, it was found that the advantages of OHDE relative to $\Lambda$CDM diminish.
Ref. \cite{Li:2024qus} found that the OHDE and the interacting HDE models are statistically as viable as the $\Lambda$CDM model for the combined datasets of DESI BAO and SN data, but these HDE models become less favored after the CMB data are included.
Ref. \cite{Tyagi:2024cqp} investigated the entropy-based approaches to HDE and the corresponding Gravity-Thermodynamics (GT) formalisms, and found that the HDE approach is statistically equivalent to the $\Lambda$CDM model for the combined datasets of PantheonPlus, DESY5, and DESI BAO.

In this paper, instead of studying an individual HDE model, we perform a comprehensive numerical study on all four categories of HDE models. Specifically, we select one or two representative models from each category, as summarized in Table \ref{tab:hde_models}. For the first category, we discuss the OHDE model. For the second category, we analyze both the RDE and GRDE models. For the third category, we examine two interacting HDE models (IHDE1 and IHDE2), one with the Hubble scale as the IR cutoff and the other with the future event horizon as IR cutoff. For the fourth category, we consider the THDE and BHDE models. Therefore, we study seven representative HDE models in this paper. In addition, we also consider the $\Lambda$CDM model as the fiducial model. For observational data, we adopt the BAO data from DESI 2024 DR1, the CMB distance priors data from Planck 2018, and the SN Ia data from PantheonPlus compilation. To perform numerical analysis, we divide these observational data into two datasets: the first dataset only includes DESI BAO and Planck 2018 CMB data, while the second dataset combines the first dataset with the PantheonPlus SN data. Moreover, we apply the $\chi^2$ statistic to perform cosmology-fits, and then calculate the Bayesian evidence of these HDE models to compare their relative performance.

%\begin{table}[h]
\begin{table}
\begin{center}
\renewcommand{\arraystretch}{1.2}
\begin{tabular}{lll}
\hline \hline
$\textbf{Model Category}$ & $\textbf{Model}$ & $\textbf{Reference}$ \\
\hline \hline
\multirow{1}{*}{HDE models with other characteristic length scale}
  &  OHDE      &  \cite{Li:2004rb}  \\
\hline
\multirow{2}{*}{HDE models with extended Hubble scale}
  &  RDE  & \cite{Gao:2007ep} \\
  &  GRDE &  \cite{Granda:2008dk}\\
\hline
\multirow{2}{*}{HDE models with dark sector interaction}
  &  IHDE1 &  \cite{Zimdahl:2007zz} \\
  &  IHDE2 &  \cite{Wang:2005jx} \\
\hline
\multirow{2}{*}{HDE models with modified black hole entropy}
  &  THDE &  \cite{Tavayef:2018xwx} \\
  &  BHDE &  \cite{Saridakis:2020zol} \\
\hline
\end{tabular}
\end{center}
\caption{HDE models and their references.}
\label{tab:hde_models}
\end{table}

The paper is organized as follows: Section \ref{sec:MoHDE} introduces seven representative HDE models. In Section \ref{sec:data_method}, we outline the observational data and the methodology used in our analysis. Section \ref{sec:results} presents the results of the joint observational constraints along with detailed discussions. In Section \ref{sec:conclusion}, we draw conclusions based on the constraints results. Finally, Section \ref{sec:Summary} provides a brief summary of the study.

\section{Holographic Dark Energy Models}
\label{sec:MoHDE}
In this section, we present the theoretical foundations of the seven representative HDE models considered in this paper.

As a comparison, we also study the standard cosmological model (i.e. $\Lambda$CDM model) as the fiducial model. The spatially flat Friedmann equation for this model can be expressed as
\begin{equation}
    E(z) \equiv \frac{H(z)}{H_0} = \sqrt{ \Omega_{r}(1+z)^4 +\Omega_{m}(1+z)^3 + \Omega_{de} },
\end{equation}
where $H(z)$ is the Hubble parameter, and $H_0=100h$ km $\text{s}^{-1} \text{Mpc}^{-1}$ is the present-day Hubble constant with $h$ being its dimensionless form. $\Omega_r$ and $\Omega_m$ denote the fractional energy densities of radiation and matter, respectively, while $\Omega_{de}$ represents the DE component and is given by $\Omega_{de}=1-\Omega_{r} -\Omega_{m}$. In this model, the EoS of DE is constant, $w=-1$, indicating that DE does not evolve over time. 

After the brief introduction of the standard model, we turn our attention to the HDE models. 
In the following subsections, we shall introduce seven representative HDE models, each illustrating a distinct theoretical construction or cutoff choice within the HDE framework.

\subsection{HDE models with other characteristic length scale}
This type of HDE model chooses the other characteristic length scale, which has nothing to do with the Hubble scale, as the IR cutoff.

\subsubsection{original holographic dark energy (OHDE) model}

In the OHDE model \cite{Li:2004rb}, an accelerating expanding universe is achieved by choosing the future event horizon as the characteristic length, defined as
\begin{equation}
    L = a\int^\infty_t \frac{d t'}{a} = a \int^\infty_a \frac{d a'}{H a'^2},
\end{equation}
where $a$ is the scale factor $a=(1+z)^{-1}$.
In this case, the Friedmann equation reads
\begin{equation}
\label{Fdmeq}
    3M_P^2 H^2 = \rho_r +\rho_{m} +\rho_{de},
\end{equation}
or equivalently,
\begin{equation}
\label{ez0}
    E(z)= \sqrt{ \frac{\Omega_r (1+z)^4 +\Omega_{m}(1+z)^3 }{1-\Omega_{de}(z)} }.
\end{equation}
Taking derivative of $\Omega_{de}$ as well, the dynamical evolution equation of $\Omega_{de}(z)$ is obtained as
\begin{equation}
\label{domdedz}
    \frac{d \Omega_{de}(z)}{d z} = -\frac{2\Omega_{de}(z)(1-\Omega_{de}(z))}{1+z} \Big( \frac{1}{2} + \frac{\sqrt{\Omega_{de}(z)}}{C} +\frac{\Omega_{r}(z)}{2(1- \Omega_{de}(z))} \Big) .
\end{equation}
Solving Eq. (\ref{domdedz}) numerically and substituting the corresponding results into Eq. (\ref{ez0}), one can obtain the redshift evolution of Hubble parameter $H(z)$ of the OHDE model, enabling cosmological constraints to be obtained.

With conservation equation
\begin{equation}
\label{conserv.eq}
    \dot{\rho}_{de}+3H\rho_{de}(1+w)=0,
\end{equation}
and taking derivative for Eq. (\ref{rhode}) with respect to $x\equiv\ln a$, the EoS is given by
\begin{equation}
    w=-\frac{1}{3} - \frac{2\sqrt{\Omega_{de}}}{3C}.
\end{equation}
It is obvious that the EoS of the OHDE evolves dynamically and satisfies $-(1+2/C)/3 \leq w \leq -1/3$ due to $0 \leq \Omega_{de} \leq 1$.
In the late-time universe with $\Omega_{de} \simeq 1$, 
if $C= 1$, $w = -1$, then HDE will be close to the cosmological constant; 
if $C > 1$, $w > -1$, then HDE will be a quintessence DE; 
if $C < 1$, $w < -1$ thus HDE will be a phantom DE.
For more discussions on this model, we refer the reader to Ref. \cite{Wang:2016och, Wang:2023gov}.

\subsection{HDE models with extended Hubble scale}

In these HDE models, IR cutoff is identified with a combination of Hubble scale and it's time derivative.

\subsubsection{Ricci dark energy (RDE) model}
The Ricci dark energy (RDE) model \cite{Gao:2007ep} is a representative model to describe DE by using the Ricci scalar curvature as the IR cutoff. In a flat universe, the Ricci scalar curvature is given by
\begin{equation}
    R= -6(2H^2 + \dot H),
\end{equation}
where the dot represents the derivative with respect to time. Taking the Ricci scalar curvature as the IR cutoff, the energy density of RDE can be expressed as
\begin{equation}
    \rho_{de} = 3\alpha M_P^2 (2H^2 + \dot H),
\end{equation}
where $\alpha$ is a constant to be determined. The Friedmann equation for this model is
\begin{equation}
    H^2= \frac{1}{3 M_P^2} ( \rho_{m}e^{-3x} +\rho_{r}e^{-4x} ) + \alpha \left( 2H^2 + \frac{1}{2}\frac{d H^2}{d x} \right).
\end{equation}
This equation can be solved as
\begin{equation}
\label{eq.rde_ez}
    E^2(z) = \Omega_{r}(1+z)^4 + \Omega_{m}(1+z)^3 + \frac{\alpha}{2-\alpha}\Omega_{m}(1+z)^3  +f_0 (1+z)^{4-\frac{2}{\alpha}},
\end{equation}
where $f_0$ is the integration constant, which can be fixed by initial condition $E_0=E(t_0)=1$:
\begin{equation}
    f_0 = 1-\Omega_r -\frac{2\Omega_m}{2-\alpha}.
\end{equation}
From Eq.(\ref{eq.rde_ez}), one can identify the RDE density
\begin{equation}
    \Omega_{de} = \frac{\alpha}{2-\alpha}\Omega_{m}(1+z)^3  +f_0 (1+z)^{4-\frac{2}{\alpha}}.
\end{equation}

The EoS of RDE satisfies
\begin{equation}
    w=-1+\frac{1+z}{3} \frac{d \ln \Omega_{de}}{dz}
\end{equation}
If $\alpha=1/2$, the RDE behaves as a cosmological constant plus a component of "dark matter".
When $1/2 \leq \alpha < 1$, the RDE has EoS $-1 \leq w \leq -1/3$. When $\alpha < 1/2$, the RDE will start from quintessence-like and evolve to phantom-like.

\subsubsection{Generalized Ricci dark energy (GRDE) model}
Granda and Oliveros proposed a generalization of the Ricci scalar as the IR cutoff in HDE models \cite{Granda:2008dk}. In this model, the cutoff is expressed as
\begin{equation}
\label{gen-ricci}
    L^{-2} = \lambda H^2 + \beta \dot H,
\end{equation}
where $\lambda$ and $\beta$ are independent model parameters. This cutoff is commonly known as the Granda-Oliver (GO) cutoff. Since the model takes a generalized Ricci scalar as the cutoff, we refer to it as the GRDE model. The corresponding holographic dark energy density is given by
\begin{equation}
    \rho_{de} = 3 M_P^2(\lambda H^2 +\beta \dot H).
\end{equation}
The Friedmann equation for this model is
\begin{equation}
\begin{split}
    E(z)^2 = & \Omega_r (1+z)^4 + \Omega_m (1+z)^3 + \frac{2\beta - \lambda}{\lambda - 2\beta - 1} \Omega_r (1+z)^4 \\
    &+ \frac{3\beta -2\lambda}{2\lambda - 3\beta - 2} \Omega_m (1+z)^3 + g_0 (1+z)^{2(\lambda -1)/ \beta},
\end{split}
\end{equation}
where the last three terms give the scale dark energy density $\Omega_{de}$, $g_0$ can be determined by initial condition $E_0=E(t_0)=1$ as
\begin{equation}
    g_0 = 1 - \frac{2\Omega_m}{3\beta -2 \alpha +2} - \frac{\Omega_r}{2\beta -\alpha +1 }.
\end{equation}

With conservation equation, the EoS is given by
\begin{equation}
    w=\frac{2(\lambda-1)+(2-3\beta)\Omega_m +(2-4\beta)\Omega_r}{3\beta (1-\Omega_m -\Omega_r)}-1.
\end{equation}
Neglecting the radiation term, the GRDE model reduces to a cosmological constant when $\lambda =1$, $\beta = 2/3$. In the late-time universe, DE becomes the dominant component. Therefore, neglecting matter and radiation, if $0<\lambda -1< \beta$, the GRDE will be a quintessence DE; if $\lambda >1$ with $\beta <0$ or $\lambda <1$ with $\beta >0$, the GRDE will be a phantom DE. For more discussions on this model, we refer the reader to Ref. \cite{Granda:2008dk}.

\subsection{HDE models with dark sector interaction}
In these HDE models, dark energy and dark matter no longer evolve independently but interact with each other. We focus on two cases. In the first case, the Hubble horizon is adopted as the IR cutoff, referred to as the IHDE1 model. In the second case, the future event horizon is used as the IR cutoff, referred to as IHDE2 model.

\subsubsection{Interacting holographic dark energy (IHDE1) model with Hubble horizon}
Taking into account the mutual interaction, the energy densities of DE and dark matter evolve according to the equations below \cite{Aboubrahim:2024spa, Aboubrahim:2024cyk}:
\begin{equation}
\label{Q1}
    \dot \rho_{m} + 3H\rho_{m}=Q,
\end{equation}
\begin{equation}
    \label{Q2}
    \dot \rho_{de} + 3H(1+w)\rho_{de}=-Q,
\end{equation}
where $Q$ denotes the interaction term. As shown in \cite{Pavon:2005yx}, the presence of interaction $Q$ enables the Hubble scale to serve as the cutoff length.

By adopting the Hubble scale as the IR cutoff and following the growth assumption in \cite{Zimdahl:2007zz}, the interaction term is given by
\begin{equation}
\label{IHDE1-Q}
    Q= 3\eta \rho_{m} H a ^\varepsilon,
\end{equation}
where $\varepsilon$ and $\eta$ are positive-definite parameters. Using the ansatz (\ref{IHDE1-Q}), the scaled Hubble rate is expressed as \cite{Zimdahl:2007zz}
\begin{equation}
\label{IHDE1-Hz}
    \frac{H(z)}{H_0} =  (1+z)^{3/2} \exp \left[ \frac{3\eta}{2\varepsilon} ( (1+z)^{-\varepsilon} -1) \right].
\end{equation}
Comparing Eq. (\ref{IHDE1-Hz}) with the corresponding quantity of the $\Lambda$CDM:
\begin{equation}
\label{ihde1-lcdm}
\begin{split}
    \frac{H(z)_{\Lambda CDM}}{H_0} &= \sqrt{\frac{\Omega_\Lambda}{\Omega_\Lambda + \Omega_m}} \left[ 1+\frac{\Omega_m}{\Omega_\Lambda} (1+z)^3 \right]^{1/2} \\
    &= 1 + \frac{3}{2}\frac{\Omega_m}{\Omega_\Lambda + \Omega_m}z + \mathcal{O}(z^2),
\end{split}
\end{equation}
we obtain the scaled Hubble rate for this IHDE1 model as follows
\begin{equation}
\label{ihde1-hz2}
\begin{split}
    E(z) &= (1+z)^{3/2} \exp \left[ \frac{3\eta }{2\varepsilon } ( (1+z)^{-\varepsilon} -1) \right] \\
    & = 1+ \frac{3}{2} (1- \eta) z + \mathcal{O}(z^2).
\end{split}
\end{equation}
As shown in \cite{Zimdahl:2007zz}, Eq. (\ref{ihde1-hz2}) must coincide with Eq. (\ref{ihde1-lcdm}) up to linear order in $z$. This requirement allows the parameter $\eta$ to be determined as
\begin{equation}
    \eta = \frac{\Omega_{de}}{\Omega_{de}+\Omega_m}.
\end{equation}
Once $\eta$ is determined, the scaled Hubble rate can be written as
\begin{equation}
    E(z) = (1+z)^{3/2} \exp \left[ \frac{3(1-\Omega_m) }{2\varepsilon } ( (1+z)^{-\varepsilon} -1) \right].
\end{equation}
In this model, the EoS is given by
\begin{equation}
    w = -\eta a^\varepsilon.
\end{equation}
Notice that DE is negligible in the early universe, it means that $\varepsilon >0$ must be satisfied. For the case of $\varepsilon >0$, the IHDE1 is a quintessence DE at present day. In the late-time universe with $\Omega_{de} \simeq 1$, the IHDE1 behaves as a cosmological constant. For more discussions on this model, we refer the reader to Ref. \cite{Zimdahl:2007zz}.

\subsubsection{Interacting holographic dark energy (IHDE2) model with future event horizon}
An alternative approach to the Hubble scale is to use the future event horizon as the IR cutoff. In this model, the interaction term takes the form $Q=3b^2 M_P^2 H (\rho_{de} +\rho_m)$ with $b^2$ the coupling constant \cite{Wang:2005jx}. By combining the interaction term and energy conservation equations (\ref{Q1}, \ref{Q2}), the evolution equation for $\Omega_{de}(z)$ is written as
\begin{equation}
    \frac{d \Omega_{de}}{d z}= -\frac{\Omega_{de}^2}{1+z} (1-\Omega_{de})\left[ \frac{1}{\Omega_{de}} +\frac{2}{C\sqrt{\Omega_{de}}} -\frac{3b^2 - \Omega_{r}}{\Omega_{de}(1-\Omega_{de})} \right].
\end{equation}
The Friedmann equation in this model satisfies
\begin{equation}
    \frac{1}{E(z)} \frac{dE(z)}{dz}= -\frac{\Omega_{de}}{1+z} (\frac{1}{2} +\frac{\sqrt{\Omega_{de}}}{C} +\frac{3b^2 -3-\Omega_r}{2\Omega_{de}} ).
\end{equation}
By numerically solving the above equation, we then obtain the evolutions of both $\Omega_{de}$ and $E(z)$ as functions of redshift.

The EoS is given by
\begin{equation}
    w=-\frac{1}{3} - \frac{2\sqrt{\Omega_{de}}}{3C} -\frac{b^2}{\Omega_{de}}.
\end{equation}
If $b^2 <0$, it means a transfer of energy from dark matter to DE,
if $b^2 >0$, it means a transfer of energy from DE to dark matter,
if $b^2 =0$, the EoS reduces to the OHDE model. In the late-time universe with $\Omega_{de} \simeq 1$, if we neglect the interaction, the behaviors of IHDE2 will be the same as OHDE. For more discussions on this model, we refer the reader to Ref. \cite{Wang:2005jx}.

\subsection{HDE models with modified black hole entropy}

In these HDE models, the HDE density depends on the specific entropy-area relationship $S \sim A \sim L^2$ of black holes, where $A=4\pi L^2$ represents the area of the horizon.

\subsubsection{Tsallis holographic dark energy (THDE) model with Hubble horizon}
Tsallis and Cirto proposed that the traditional Boltzmann-Gibbs additive entropy should be generalized to a non-additive entropy, known as Tsallis entropy \cite{Tsallis:2012js}. This generalized entropy is given by
\begin{equation}
\label{S_Tsallis}
    S_\delta =\gamma A^\delta,
\end{equation}
where $\gamma$ is an unknown constant, and $\delta$ denotes the non-additivity parameter. Building on this concept, M. Tavayef et al. introduced the Tsallis HDE (THDE) model \cite{Tavayef:2018xwx}. By substituting the relation (\ref{S_Tsallis}) into the ultraviolet (UV) and IR (L) cutoff $ \rho_{de} \leq S L^{-4}$, the energy density of DE is modified as
\begin{equation}
    \rho_{de} = B L^{2\delta -4},
\end{equation}
where $B = 3C^2M_P^2$. In this model, the Hubble scale is considered a suitable candidate for the IR cutoff, which can lead to the late-time accelerated expansion of the universe. The Friedmann equation of the THDE model also satisfies
\begin{equation}
\label{Fmeq}
    E(z)= \sqrt{ \frac{ \Omega_r(1+z)^4 + \Omega_{m}(1+z)^3 }{1-\Omega_{de}(z)} }.
\end{equation}
The evolution equation of $\Omega_{de}$ is governed by the equation
\begin{equation}
    \frac{d \Omega_{de}}{d z}= - \frac{3(\delta -1)}{1+z} \Omega_{de} \left( \frac{1-\Omega_{de} - 5\Omega_{r}}{1-(2-\delta)\Omega_{de}} \right).
\end{equation}

With conservation equation (\ref{conserv.eq}), the EoS is given by
\begin{equation}
    w=\frac{\delta-1}{(2-\delta)\Omega_{de}-1}.
\end{equation}
When $\delta =2$, the EoS of this model coincides with that of $\Lambda$CDM, where $w=-1$. 
Assuming the present-day $\Omega_{de}=0.7$, 
if $\delta >2$, the THDE will be a quintessence DE,
if $\delta <2$, the THDE will be a phantom DE.
For more discussions on this model, we refer the reader to Ref. \cite{Tavayef:2018xwx}.

\subsubsection{Barrow holographic dark energy (BHDE) model with future event horizon}
In Ref. \cite{Barrow:2020tzx, Barrow:2020kug}, it was suggested that quantum-gravitational effects may cause deformations on the black hole surface, leading to deviations from the standard Bekenstein-Hawking entropy. This modified entropy, known as Barrow entropy, is expressed as
\begin{equation}
    S_B=\left( \frac{A}{A_0} \right)^{1+\Delta/2},
\end{equation}
where $A$ is the standard horizon area, $A_0$ is the Planck area, and $\Delta$ is a parameter quantifies quantum-gravitational deformation. Here, $0 \leq \Delta \leq 1$, with $\Delta = 0$ corresponding to a smooth spacetime structure and $\Delta = 1$ representing the most intricate deformation. By applying Barrow entropy to UV/IR relation, E. N. Saridakis proposed the Barrow HDE (BHDE) model \cite{Saridakis:2020zol}. In this model, the HDE density is modified as
\begin{equation}
    \rho_{de}=B L^{\Delta-2}.
\end{equation}
In the BHDE model, the future event horizon serves as the IR cutoff. The Friedmann equation also satisfies Eq. (\ref{Fmeq}), and the evolution equation for $\Omega_{de}$ is \cite{Denkiewicz:2023hyj}
\begin{equation}
\begin{split}
    \frac{d \Omega_{de}}{d z} = &-\frac{\Omega_{de}(1-\Omega_{de})}{1+z} [ (1+\frac{\Delta}{2})\mathcal{F}_r + (\Delta +1)\mathcal{F}_m \\
    &+G(1-\Omega_{de})^{\frac{\Delta}{2(\Delta -2)}} (\Omega_{de})^{\frac{1}{2-\Delta}} ],
\end{split}
\end{equation}
where
\begin{equation}
\begin{split}
    \mathcal{F}_r &=\frac{2\Omega_r (1+z)^4}{\Omega_m (1+z)^3 + \Omega_r (1+z)^4},\\
    \mathcal{F}_m &=\frac{\Omega_m (1+z)^3}{\Omega_m (1+z)^3 + \Omega_r (1+z)^4},\\
    G &\equiv (2-\Delta) ( \frac{B}{3M_P^2} )^{\frac{1}{\Delta -2}} ( H_0 \sqrt{\Omega_m (1+z)^3 + \Omega_r (1+z)^4 })^{\frac{\Delta}{2-\Delta}}.
\end{split}
\end{equation}
The EoS is given by
\begin{equation}
    w=-\frac{1+\Delta}{3}-\frac{G}{3}(\Omega_{de})^{\frac{1}{2-\Delta}} (1-\Omega_{de})^{\frac{\Delta}{2(\Delta-2)}} e^{\frac{3\Delta}{2(2-\Delta)}x}.
\end{equation}
In the standard case of $\Delta =0$, the EoS reduces to that of the OHDE model. 
For $\Delta \lesssim 0.5$, the BHDE behaves as a quintessence DE,
for $\Delta >0.5$, the BHDE behaves as a quintom DE. 
In the late-time universe with $\Omega_{de} \simeq 1$, if $\Delta =0$, the behaviors of BHDE will be the same as OHDE.
For more discussions on this model, we refer the reader to Ref. \cite{Saridakis:2020zol}.

\section{Data and Methodology}
\label{sec:data_method}

\subsection{Data}
\subsubsection{Baryon Acoustic Oscillation}

For the BAO data, we adopt the first-year data released by the DESI collaboration \cite{DESI:2024mwx}, which includes observations from four different classes of extragalactic targets: the bright galaxy sample (BGS) \cite{Hahn:2022dnf}, luminous red galaxies (LRG) \cite{DESI:2022gle}, emission line galaxies (ELG) \cite{Raichoor:2022jab}, and quasars (QSO)  \cite{Chaussidon:2022pqg}. Table \ref{tab:BAO} presents the tracers, effective redshifts, observables, and measurement values for the seven BAO data points.

\begin{table}
    \centering
    \begin{tabular}{c|c|c|c|c}
    \hline \hline
    \textbf{Tracer} & \textbf{$z_{\text{eff}}$} & \textbf{$D_M / r_d$} & \textbf{$D_H / r_d$} & \textbf{$r$ or $D_V / r_d$} \\
    \hline \hline
    BGS & 0.295 & --- & --- & 7.93 $\pm$ 0.15 \\
    LRG1 & 0.510 & 13.62 $\pm$ 0.25 & 20.98 $\pm$ 0.61 & $-0.445$ \\
    LRG2 & 0.706 & 16.85 $\pm$ 0.32 & 20.08 $\pm$ 0.60 & $-0.420$ \\
    LRG3+ELG1 & 0.930 & 21.71 $\pm$ 0.28 & 17.88 $\pm$ 0.35 & $-0.389$ \\
    ELG2 & 1.317 & 27.79 $\pm$ 0.69 & 13.82 $\pm$ 0.42 & $-0.444$ \\
    QSO & 1.491 & --- & --- & 26.07 $\pm$ 0.67 \\
    Lya QSO & 2.330 & 39.71 $\pm$ 0.94 & 8.52 $\pm$ 0.17 & $-0.477$ \\
    \hline
    \end{tabular}
    \caption{Statistics from the DESI DR1 BAO measurements. Note that for each sample DESI DR1 measures either both $D_M / r_d$ and $D_H / r_d$, which are correlated with a coefficient $r$, or $D_V / r_d$.}
    \label{tab:BAO}
\end{table}

The quantities of BAO listed in Table \ref{tab:BAO} correspond to several key distances: $D_M$, $D_H$, and $D_V$. In a spatially flat FLRW universe, the transverse comoving distance $D_M$ at redshift $z$ is defined as \cite{DESI:2024mwx}:
\begin{equation}
\label{codist}
    D_M(z) = \frac{c}{H_0} \int_0^z \frac{d z'}{H(z') /H_0},
\end{equation}
where $c$ is the speed of light. The distance variable $D_H$ is related to the Hubble parameter $H(z)$ as $D_H(z) = c/H(z)$. The angle-averaged distance $D_V$ is given by $D_V(z) = [zD_M(z)^2 D_H(z) ]^{1/3}$.

In Table \ref{tab:BAO}, BAO measurements depend on the radius of the sound horizon at the drag epoch $r_d$. This represents the distance that sound can travel between the Big Bang and the drag epoch, which marks the time when baryons decoupled. The sound horizon can be expressed as \cite{DESI:2024mwx}:
\begin{equation}
\label{rs}
    r_s(z) = \int_{z}^\infty \frac{c_s(z')}{H(z')}dz',
\end{equation}
where $c_s(z)$ is the speed of sound which, prior to recombination, is given by
\begin{equation}
    c_s(z)=\frac{c}{\sqrt{3\left( 1+\frac{3\rho_b}{4\rho_r} \right)}}
\end{equation}
where $\rho_b$ and $\rho_r$ are the baryon and radiation densities, respectively. Specifically, $\bar{R_b}/(1+z) = 3\rho_b/(4\rho_r)$, and $\bar{R_b} =31500 \Omega_b h^2 (T_{\text{CMB}} /2.7\text{K})^{-4}$, where $\Omega_b$ is the fractional density of baryons. The radiation term in the $H(z)$ can be determined by the matter-radiation equality relation $\Omega_r = \Omega_m/(1+z_{eq})$, and $z_{eq} = 2.5\times 10^4 \Omega_m h^2 (T_{\text{CMB}} /2.7\text{K})^{-4}$. We assume the CMB temperature to be $T_{\text{CMB}}=2.7255$K.

In practice, the redshift of the drag epoch, $z_d$, is approximated by \cite{Eisenstein:1997ik}:
\begin{align}
    z_d &= \frac{1291(\Omega_m h^2)^{0.251}}{1+0.659(\Omega_m h^2)^{0.828}} [1+b_1 (\Omega_b h^2)^{b_2}],\\
    b_1 &= 0.313 (\Omega_m h^2)^{-0.419} [1+0.607(\Omega_m h^2)^{0.674}],\\
    b_2 &= 0.238(\Omega_m h^2)^{0.223}.
\end{align}
Hence, the sound horizon at the drag epoch is $r_d = r_s(z_d)$. The data vector $D$ can be constructed as
\begin{equation}
\label{vetorD}
    D \equiv \begin{pmatrix}
    D_M / r_d \\
    D_H / r_d \\
\end{pmatrix},
\end{equation}
with its covariance matrix defined as \cite{strang2000introduction}:
\begin{equation}
    \text{Cov}_{BAO} = \begin{bmatrix}
    \sigma_1^2 & r \cdot \sigma_1 \cdot \sigma_2 \\
    r \cdot \sigma_1 \cdot \sigma_2 & \sigma_2^2
    \end{bmatrix},
\end{equation}
where $\sigma_1$ and $\sigma_2$ denote the standard deviations of $D_M / r_d$ and $D_H / r_d$, respectively. The correlation coefficient between $D_M / r_d$ and $D_M / r_d$, denoted as $r$, is provided in Table \ref{tab:BAO}.

\subsubsection{Cosmic Microwave Background}
For the CMB data, we use the distance priors from Planck 2018 release \cite{Planck:2018vyg}. The method of distance priors \cite{Efstathiou:1998xx, Wang:2006ts, Wang:2007mza, Chen:2018dbv, Zhai:2019nad} serves as a compressed dataset that encapsulates key information from the full CMB data. This approach allows us to substitute the full CMB power spectrum with a more compact representation while retaining key cosmological information.

The distance priors contain two primary features of the CMB power spectrum: the shift parameter $R$ and the acoustic scale $l_a$. The shift parameter $R$ affects the peak heights in the CMB temperature power spectrum along the line of sight, while the acoustic scale $l_a$ influences the spacing of the peaks in the transverse direction. These parameters are defined as
\begin{align}
    R &\equiv \frac{ D_M(z_*)\sqrt{\Omega_m H_0^2}}{c},\\
    l_a &\equiv \frac{\pi D_M(z_*)}{r_s(z_*)},
\end{align}
where $z_*$ is the redshift at the photon decoupling epoch, which can be calculated by an approximate formula \cite{Hu:1995en}:
\begin{equation}
    z_* = 1048[1+0.00124(\Omega_b h^2)^{-0.738}][1+g_1(\Omega_m h^2)^{g_2}],
\end{equation}
where
\begin{align}
    g_1 &= \frac{0.0783(\Omega_b h^2)^{-0.238}}{1+39.5(\Omega_b h^2)^{0.763}},\\
    g_2 &= \frac{0.560}{1+21.1(\Omega_b h^2)^{1.81}}.
\end{align}
As pointed out by Ref. \cite{Wang:2007mza}, the baryon density $\Omega_b h^2$ should be included as an estimated parameter in the data analysis.
This is because its value is essential for computing the redshift $z_*$ at the photon decoupling epoch. 
Therefore, the CMB shift parameters $R$ and $l_a$ inherit an explicit dependence on $\Omega_b h^2$ through redshift $z_*$.

Since $\Omega_b h^2$ is correlated with $R$ and $l_a$, we should take into account their covariance. The data vector and its covariance matrix are the following respectively \cite{Zhai:2018vmm}:
\begin{equation}
\label{vetorV}
    V^{data} \equiv \begin{pmatrix}
    R \\
    l_a \\
    \Omega_b h^2 \\
    \end{pmatrix}
    =
    \begin{pmatrix}
    1.74963 \\
    301.80845 \\
    0.02237 \\
    \end{pmatrix},
\end{equation}

\begin{equation}
    \text{Cov}_{CMB} = 10^{-8} \times \\
    \begin{bmatrix}
    1598.9554 & 17112.007 & -36.311179 \\
    17112.007 &  811208.45 &  -494.79813 \\
    -36.311179 &  -494.79813 &  2.1242182
    \end{bmatrix}.
\end{equation}

\subsubsection{Type Ia Supernovae}
For the SN data, we use a subset of the PantheonPlus compilation \cite{Brout:2022vxf}, consisting of 1701 data points. We remove all SN with $z < 0.01$, as these data points are influenced by model-dependent peculiar velocities that need to be considered separately. This cut leaves us with 1590 data points, spanning a redshift range $0.01016 \leq z \leq 2.26137$. The covariance matrix, $\text{Cov}_{SN}$, includes both statistical and systematic errors.

The theoretical distance modulus $\mu$ in a flat universe is given by
\begin{equation}
    \mu_{th} = 5 \log_{10} \left[ \frac{d_L(z_{hel}, z_{cmb})}{\text{Mpc}} \right] +25,
\end{equation}
where $z_{hel}$ and $z_{cmb}$ are the heliocentric and CMB rest-frame redshifts of SN. The luminosity distance $d_L$ is
\begin{equation}
    d_L(z_{hel}, z_{cmb})= (1+z_{hel}) r(z_{cmb}),
\end{equation}
where $r(z)$ is the comoving distance given by Eq. (\ref{codist}). The observation of distance modulus $\mu$ is given by the following empirical linear relation:
\begin{equation}
    \mu_{obs} = m_B - M_B + \alpha X_1 - \beta \mathcal{C} - \delta_{bias} + \delta_{host},
\end{equation}
where $m_B$ is the observed peak magnitude in the rest-frame of the B band, $M_B$ is the fiducial magnitude of a SN, $X_1$ describes the time stretching of light-curve, and $\mathcal{C}$ describes the supernova color at maximum brightness, $\delta_{bias}$ is a correction term to account for selection biases, and $\delta_{host}$ is the luminosity correction for residual correlations. Note that $\alpha$ and $\beta$ are global nuisance parameters relating stretch and color, respectively.

\subsection{Methodology}
\subsubsection{chi-squared $\chi^2$ statistic}
The $\chi^2$ statistic quantifies the goodness of fit between predicted values from cosmological models and actual measurements from astronomical observations. By minimizing the $\chi^2$ function, one can identify the model parameters that best describe the observed universe.

There are two methods to calculate the $\chi^2$ function. For independent data points, the $\chi^2$ function is defined as
\begin{equation}
    \chi^2_\xi = \frac{(\xi_{th} - \xi_{obs})^2}{\sigma_\xi ^2},
\end{equation}
where $\xi_{th}$ is the theoretically predicted value, $\xi_{obs}$ is the experimentally measured value, and $\sigma_\xi$ is the standard deviation. For correlated data points, the $\chi^2$ function is given by
\begin{equation}
    \chi^2 = \Delta \xi ^T \text{Cov}^{-1} \Delta \xi,
\end{equation}
where $\Delta \xi \equiv \xi_{th} - \xi_{obs} $, and Cov is a covariance matrix that characterizes the errors in the data.

For $\textbf{BAO data}$, the $\chi^2$ function is split into two parts:
\begin{equation}
    \chi^2_{BAO} =  \chi^2_1 + \chi^2_2,
\end{equation}
where $\chi^2_1$ represents the data $D_V / r_d$ from tracer BGS and QSO, and is expressed as
\begin{equation}
    \chi^2_1 = \sum_i \frac{(\xi_i^{th}-\xi_i^{data})^2}{\sigma_i^2}.
\end{equation}
The second term, $\chi^2_2$, represents the data for $D_M / r_d$ and $D_H / r_d$ from tracers LRG1, LRG2, LRG3+ELG1, ELG2, and Lya QSO, and is expressed as
\begin{equation}
    \chi^2_2 = \sum_i \Delta \text{D}_i^T  \text{Cov}_{BAO}^{-1} \Delta \text{D}_i,
\end{equation}
where $\Delta \text{D}_i = D_i^{th} - D_i^{data}$ is the data vector constructed by Eq. (\ref{vetorD}).

For $\textbf{CMB data}$, the $\chi^2$ function for the CMB distance priors can be expressed as
\begin{equation}
    \chi^2_{CMB} =  \Delta \text{V}^T  \text{Cov}_{CMB}^{-1} \Delta \text{V},
\end{equation}
where $\Delta \text{V} = V^{th} - V^{data}$ is the data vector constructed by Eq. (\ref{vetorV}).

For $\textbf{SN data}$, the $\chi^2$ can be calculated as
\begin{equation}
    \chi^2_{SN} =  \Delta \mu^T \cdot \text{Cov}_{SN}^{-1} \cdot \Delta \mu,
\end{equation}
where $\Delta \mu \equiv \mu_{th} - \mu_{obs}$ is the data vector, and $\text{Cov}_{SN}^{-1}$ is the inverse matrix of the total covariance matrix provided by the Supernova Collaboration \cite{Brout:2022vxf}.

Since we use the BAO data from DESI DR1, the CMB data from Planck 2018 distance priors, and the SN data from PantheonPlus, the total $\chi^2$ is
\begin{equation}
    \chi^2 = \chi^2_{BAO} + \chi^2_{CMB} + \chi^2_{SN}.
\end{equation}

In this work, we sample from the dark energy parameter posterior distributions using the MCMC code $\mathbf{Cobaya}$ \cite{Torrado:2020dgo} with the $\mathbf{mcmc}$ sampler \cite{Lewis:2013hha, Lewis:2002ah}. Convergence of an MCMC run is assessed using the Gelman-Rubin statistic \cite{Gelman:1992zz} with a tolerance of $|R-1| < 0.01$. We analyze the MCMC chains using $\mathbf{Getdist}$  \cite{Lewis:2019xzd} to visualize the contour plots for the resulting posterior distributions. 

In Table \ref{tab:flatprior}, we present the flat prior ranges on which the parameters are left to freely vary. To ensure a relatively fair comparison between models, we impose a uniform prior width on all additional parameters that beyond the $\Lambda$CDM.

%\begin{table}[h]
\begin{table}
\begin{center}
\renewcommand{\arraystretch}{1.2}
\begin{tabular}{lll}
\hline \hline
$\textbf{Model}$ & $\textbf{Parameters}$ & $\textbf{Priors}$ \\
\hline \hline
\multirow{3}{*}{$\Lambda$CDM}
  &  $\Omega_m$      & $\mathcal{U}[0.01, 0.99]$ \\
  &  $\Omega_{b}h^2$ & $\mathcal{U}[0.005, 0.1]$ \\
  &  $h$         & $\mathcal{U}[0.2, 1.0]$ \\
\hline
\multirow{1}{*}{OHDE}
  &  $C$         & $\mathcal{U}[0.01, 2]$ \\
\hline
\multirow{1}{*}{RDE}
  &  $\alpha$    & $\mathcal{U}[0.01, 2]$ \\
\hline
\multirow{2}{*}{GRDE}
  &  $\lambda$   & $\mathcal{U}[0.01, 2]$ \\
  &  $\beta$     & $\mathcal{U}[0.01, 2]$ \\
\hline
\multirow{1}{*}{IHDE1}
  &  $\epsilon$  & $\mathcal{U}[0.01, 2]$ \\
\hline
\multirow{2}{*}{IHDE2}
  &  $b^2$       & $\mathcal{U}[-1.0, 1.0]$ \\
  &  $C$         & $\mathcal{U}[0.01, 2]$ \\
\hline
\multirow{1}{*}{THDE}
  &  $\delta$    & $\mathcal{U}[1.01, 3]$ \\
\hline
\multirow{1}{*}{BHDE}
  &  $\Delta$    & $\mathcal{U}[0.01, 2]$ \\
\hline
\end{tabular}
\end{center}
\caption{Parameters and priors used in the analysis. $\mathcal{U}[a, b]$ represents an uniform distribution from a to b.}
\label{tab:flatprior}
\end{table}

\subsubsection{Bayesian evidence}
Bayesian evidence, also known as the marginal likelihood, plays a pivotal role in model selection and parameter estimation within the Bayesian framework. Specifically, it quantifies how well a model fits the data by integrating the prior distribution of the model parameters with the likelihood function. Here, we use the $\mathbf{PolyChord}$ sampler \cite{Handley:2015fda, Handley:2015vkr}, a nested sampler incorporated in $\mathbf{Cobaya}$, to compute the Bayesian evidence for our models.

Given some dataset $\mathcal{D}$, a model $\mathcal{M}$ with parameter $\theta$ can be used to calculate the likelihood $\mathcal{L}_\mathcal{M}(\theta) \equiv P(\mathcal{D}|\theta,\mathcal{M})$. By applying Bayes' theorem, this can be inverted to determine the posterior distribution of the parameters:
\begin{equation}
\label{bayestheorem}
    P(\theta|\mathcal{D},\mathcal{M}) \equiv \mathcal{P}_\mathcal{M}(\theta) = \frac{\mathcal{L}_\mathcal{M}(\theta) \pi_\mathcal{M}(\theta)}{\mathcal{Z}_\mathcal{M}},
\end{equation}
where $\pi_\mathcal{M}(\theta) \equiv P(\theta|\mathcal{M})$ is the prior degree of belief on the values of the parameters and $\mathcal{Z}_\mathcal{M} \equiv P(\mathcal{D}|\mathcal{M})$ is the evidence or marginal likelihood, calculated as
\begin{equation}
    \mathcal{Z}_\mathcal{M} = \int \mathcal{L}_\mathcal{M}(\theta) \pi_\mathcal{M}(\theta) d \theta.
\end{equation}
The evidence can be neglected in model fitting, but it becomes crucial in model comparison.

When comparing two different models, $\mathcal{M}_1$ and $\mathcal{M}_2$, the ratio of their posterior probabilities, $P_1$ and $P_2$, is proportional to the ratio of their evidence. This relationship can be expressed as
\begin{equation}
    \frac{P_1(\theta_1|\mathcal{D},\mathcal{M}_1)}{P_2(\theta_2|\mathcal{D},\mathcal{M}_2)} = \frac{\pi_{\mathcal{M}_1}(\theta_1) \mathcal{Z}_{\mathcal{M}_1}}{\pi_{\mathcal{M}_2}(\theta_2) \mathcal{Z}_{\mathcal{M}_2}}.
\end{equation}
This ratio between posteriors leads to the definition of the Bayes Factor $B_{12}$, which in logarithmic scale is written as
\begin{equation}
    \ln \mathcal{B}_{12} \equiv \log \left[ \frac{\mathcal{Z}_{\mathcal{M}_1}}{\mathcal{Z}_{\mathcal{M}_2}} \right] = \ln [ \mathcal{Z}_{\mathcal{M}_1} ] - \ln [\mathcal{Z}_{\mathcal{M}_2}].
\end{equation}
If $\ln \mathcal{B}_{12}$ is larger (smaller) than unity, the data favor the model $\mathcal{M}_1$ ($\mathcal{M}_2$). In our analysis, we take the $\Lambda$CDM as the reference model $\mathcal{M}_2$.

\section{Results and Discussion}
\label{sec:results}
Following the description of the data and methodology, we present the corresponding cosmology-fits results and the corresponding discussions to each model. The best-fit values with $1 \sigma$ confidence level (CL) for each models are listed in Table \ref{tab:bfp}.

%\begin{turnpage}
\begin{table}
    \centering
    \begin{tabular}{lccccc}
    \hline \hline
    \\[-1mm]         & \multicolumn{5}{c}{DESI BAO + CMB}        \\[+1mm]
    \cline{2-6}\\[-1mm]
      \textbf{Model}   & $\Omega_m$  & $\Omega_b h^2$  &  $h$  &  Parameter 1   & Parameter 2  \\[+1mm]
     \hline \\[-1mm]
      $\Lambda$CDM     & $0.3218^{+0.0098}_{-0.0100}$ & $0.02230^{+0.00025}_{-0.00024}$  & $0.6689^{+0.0074}_{-0.0070}$ & - & - \\[+1mm]
      OHDE        & $0.2802^{+0.0348}_{-0.0358}$ & $0.02238^{+0.00028}_{-0.00030}$ & $0.7143^{+0.0533}_{-0.0423}$ & $C=0.5287^{+0.1131}_{-0.0934}$ & - \\[+1mm]
      RDE         & $0.1376^{+0.0101}_{-0.0021}$ & $0.02276^{+0.00035}_{-0.00030}$ & $0.9995^{+0.0004}_{-0.0351}$ & $\alpha=0.1911^{+0.0083}_{-0.0069}$ & - \\[+1mm]
      GRDE        & $0.2991^{+0.0381}_{-0.0436}$ & $0.02235^{+0.00036}_{-0.00031}$ & $0.6915^{+0.0563}_{-0.0416}$ & $\lambda=0.8539^{+0.1820}_{-0.1466}$ & $\beta=0.5543^{+0.1299}_{-0.1034}$ \\[+1mm]
      IHDE1       & $0.3497^{+0.0156}_{-0.0128}$ & $0.02240^{+0.0003}_{-0.0003}$ & $0.6392^{+0.0122}_{-0.0131}$ & $\varepsilon=1.8465^{+0.0618}_{-0.0533}$ & - \\[+1mm]
      IHDE2       & $0.2710^{+0.0487}_{-0.0438}$ & $0.02239^{+0.00033}_{-0.00034}$ & $0.7263^{+0.0686}_{-0.0587}$ & $b^2=-0.0018^{+0.0020}_{-0.0019}$ & $C= 0.4913^{+0.1845}_{-0.1124}$ \\[+1mm]
      THDE       & $0.3200^{+0.0237}_{-0.0245}$ & $0.02228^{+0.00030}_{-0.00029}$ & $0.6710^{+0.0281}_{-0.0245}$ & $\delta=2.0211^{+0.2430}_{-0.1732}$ & - \\[+1mm]
      BHDE        & $0.2921^{+0.0321}_{-0.0338}$ & $0.02236^{+0.00028}_{-0.00030}$ & $0.7000^{+0.0459}_{-0.0361}$ & $\Delta=0.2349^{+0.0740}_{-0.0724}$ & - \\[+1mm]
     \hline \hline \\[-1mm]
      & \multicolumn{5}{c}{DESI BAO + CMB + PantheonPlus}              \\[+1mm]
    \cline{2-6}\\[-1mm]
      \textbf{Model}   & $\Omega_m$  & $\Omega_b h^2$  &  $h$  &  Parameter 1   & Parameter 2  \\[+1mm]
     \hline \\[-1mm]
      $\Lambda$CDM        & $0.3221^{+0.0103}_{-0.0094}$ & $0.02229^{+0.00024}_{-0.00024}$ & $0.6686^{+0.0067}_{-0.0072}$ & - & - \\[+1mm]
      OHDE        & $0.3283^{+0.0154}_{-0.0141}$ & $0.02246^{+0.00030}_{-0.00028}$ & $0.6570^{+0.0134}_{-0.0142}$ & $C=0.6929^{+0.0664}_{-0.0521}$ & - \\[+1mm]
      RDE        & $0.3434^{+0.0158}_{-0.0152}$ & $0.02325^{+0.00030}_{-0.00029}$ & $0.6196^{+0.0125}_{-0.0123}$ & $\alpha=0.3003^{+0.0097}_{-0.0092}$ & - \\[+1mm]
      GRDE        & $0.3264^{+0.0159}_{-0.0170}$ & $0.02236^{+0.00035}_{-0.00033}$ & $0.6618^{+0.0181}_{-0.0154}$ & $\lambda=0.9907^{+0.0757}_{-0.0775}$ & $\beta=0.6518^{+0.0565}_{-0.0612}$ \\[+1mm]
      IHDE1        & $0.3451^{+0.0128}_{-0.0123}$ & $0.02241^{+0.00031}_{-0.00031}$ & $0.6433^{+0.0114}_{-0.0117}$ & $\varepsilon=1.8282^{+0.0530}_{-0.0473}$ & - \\[+1mm]
      IHDE2       & $0.3279^{+0.0167}_{-0.0165}$ & $0.02239^{+0.00034}_{-0.00033}$ & $0.6596^{+0.0174}_{-0.0157}$ & $b^2=-0.0002^{+0.0017}_{-0.0017}$ & $C= 0.7171^{+0.0867}_{-0.0704}$ \\[+1mm]
      THDE        & $0.3233^{+0.0164}_{-0.0138}$ & $0.02231^{+0.00029}_{-0.00029}$ & $0.6666^{+0.0153}_{-0.0151}$ & $\delta= 1.9818^{+0.1261}_{-0.1047}$ & - \\[+1mm]
      BHDE        & $0.3261^{+0.0157}_{-0.0137}$ & $0.02241^{+0.00030}_{-0.00027}$ & $0.6606^{+0.0130}_{-0.0147}$ & $\Delta=0.1581^{+0.0298}_{-0.0379}$ & - \\[+1mm]
    \hline \hline
    \end{tabular}
    \\[+1mm]
    \begin{flushleft}
    \caption{Cosmological parameters results from DESI BAO + CMB and DESI BAO + CMB + PantheonPlus datasets, in the baseline flat $\Lambda$CDM model and seven representative HDE models, as listed. The quoted values correspond to the best-fit estimates with 1$\sigma$ confidence level.}
    \label{tab:bfp}
    \end{flushleft}
\end{table}
%\end{turnpage}

\subsection{$\Lambda$CDM model}
\begin{figure}[ht]
    \centering
    \includegraphics[width=0.7\textwidth]{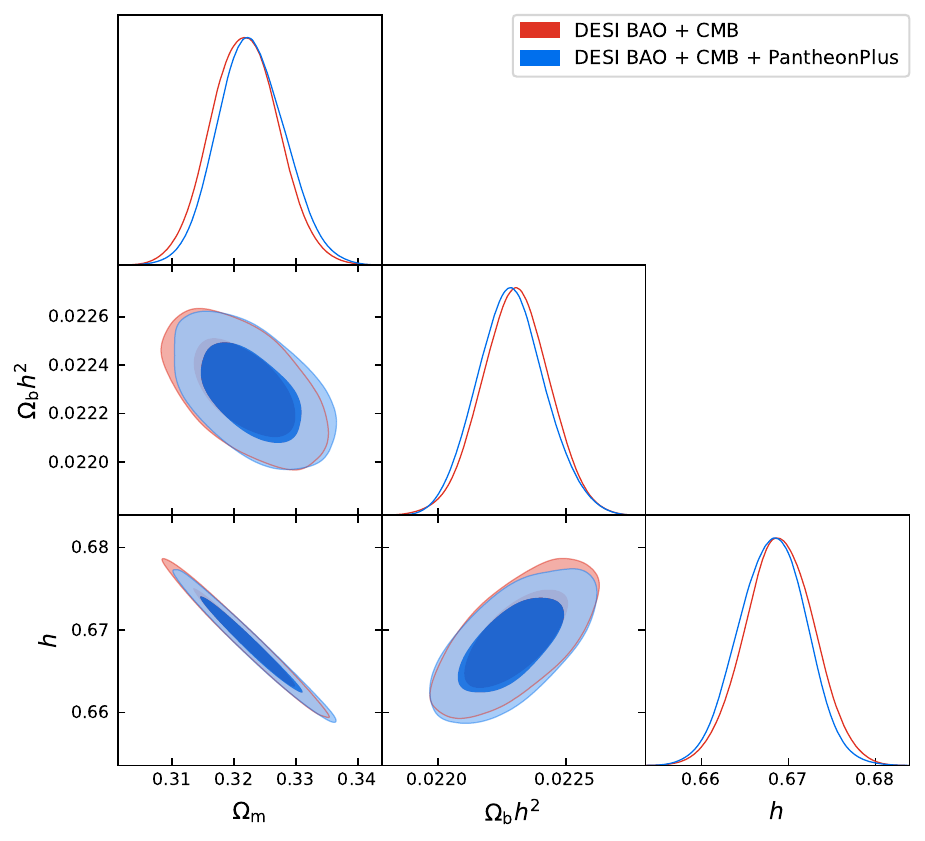}
    \caption{ One-dimensional posterior distributions and two-dimensional marginalized contours at 1$\sigma$ and 2$\sigma$ levels for $\Lambda$CDM model}
    \label{fig.LCDM}
\end{figure}

We choose the $\Lambda$CDM model as the fiducial model. Fig. \ref{fig.LCDM} presents the one-dimensional posterior distributions and two-dimensional marginalized contours of combined observational constraints for the $\Lambda$CDM model. The red contours represents constraints from the DESI BAO + CMB, while the blue contours represents constraints from the DESI BAO + CMB + PantheonPlus. Both datasets give similar results, showing good compatibility. 

\subsection{OHDE model}
The first HDE model considered in this paper is the OHDE model, which uses the future event horizon as the IR cutoff.

\begin{figure}[ht]
    \centering
    \includegraphics[width=0.8\textwidth]{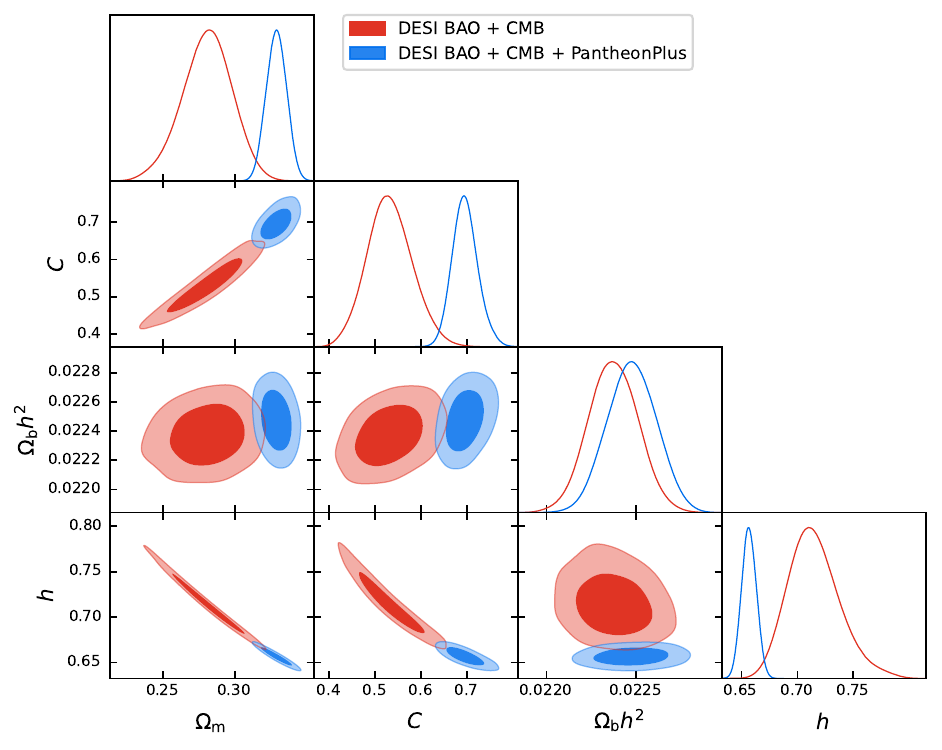}
    \caption{One-dimensional posterior distributions and two-dimensional marginalized contours at 1$\sigma$ and 2$\sigma$ levels for OHDE model.}
    \label{fig.OHDE}
\end{figure}

Fig. \ref{fig.OHDE} presents the one-dimensional posterior distributions and two-dimensional marginalized contours of combined observational constraints for the OHDE model. While the DESI BAO+CMB (red) data alone provides a broad constraint, the inclusion of PantheonPlus data (blue) generally tightens the parameter constraints, particularly for $\Omega_m$, $C$, and $h$. The inclusion of PantheonPlus data leads to higher values of $\Omega_m$ and $C$, and lower value of $h$. It must be emphasized that there is a significant discrepancy between the results for the blue contours and those for the red contours. 

The best-fit values derived from the sample likelihood constraints are listed in Table \ref{tab:bfp}. For the DESI BAO+CMB dataset, the best-fit values with $1\sigma$ confidence level (CL) are $\Omega_m=0.2802^{+0.0348}_{-0.0358}$, $h=0.7143^{+0.0533}_{-0.0423}$, and $C=0.5287^{+0.1131}_{-0.0934}$, which yields the EoS of $w=-1.40^{+0.21}_{-0.26}$. In contrast, for the DESI BAO+CMB+SN dataset, we have $\Omega_m=0.3283^{+0.0154}_{-0.0141}$, $h=0.6570^{+0.0134}_{-0.0142}$, $C=0.6929^{+0.0664}_{-0.0521}$, resulting in $w=-1.12^{+0.07}_{-0.07}$. Based on the parameter constrain results given by the two datasets, the discrepancies\footnote{Given the central values $\mu_X$ and $\mu_Y$ of quantities $X$ and $Y$, along with their asymmetric uncertainties $\sigma_{X_+}$, $\sigma_{X_-}$, and $\sigma_{Y_+}$, $\sigma_{Y_-}$, the sigma deviation is calculated as: $\text{Sigma deviation} = \Delta \mu / \sqrt{ \sigma_X^2 + \sigma_Y^2 }$, where $\Delta \mu = |\mu_X - \mu_Y|$, $\sigma_i^2 = (\sigma_{i_+} + \sigma_{i_-})/2$.} for $\Omega_m$, $C$, and $h$ are $1.25 \sigma$, $1.29 \sigma$, and $1.15 \sigma$, respectively.

Both datasets (BAO+CMB and BAO+CMB+SN) give a EoS of HDE $w<-1$, implying that the OHDE has phantom-like characteristics. Moreover, only taking into account the BAO+CMB dataset, the EoS of the OHDE deviates significantly from the cosmological constant (i.e. $w=-1$).
On the contrary, the inclusion of PantheonPlus SN data leads to the EoS of the OHDE being much closer to the cosmological constant.
This implies that under the framework of HDE with future event horizon as the IR cutoff, the PantheonPlus dataset does not favor a dynamical DE.

In fact, this phenomenon of parameter discrepancy has already been discovered in previous literature. 
Using SDSS BAO data, Planck 2018 CMB data, and Pantheon SN data, Ref. \cite{Colgain:2021beg} investigated a specific HDE model (i.e. the OHDE model) and found evidence of such discrepancy.
Ref. \cite{Colgain:2021beg} attributed this phenomenon to the presence of a turning point in $E(z)$ in the OHDE model. 
In the present work, we study seven HDE models and find that the key factor responsible for the parameter discrepancy is the choice of the characteristic length $L$. Specifically, when the future event horizon is adopted as the IR cutoff, the parameter discrepancy would emerge.

\subsection{RDE model}
The second HDE model considered in this paper is the RDE model, which uses the Ricci scalar curvature as the IR cutoff.

\begin{figure}[ht]
    \centering
    \includegraphics[width=0.8\textwidth]{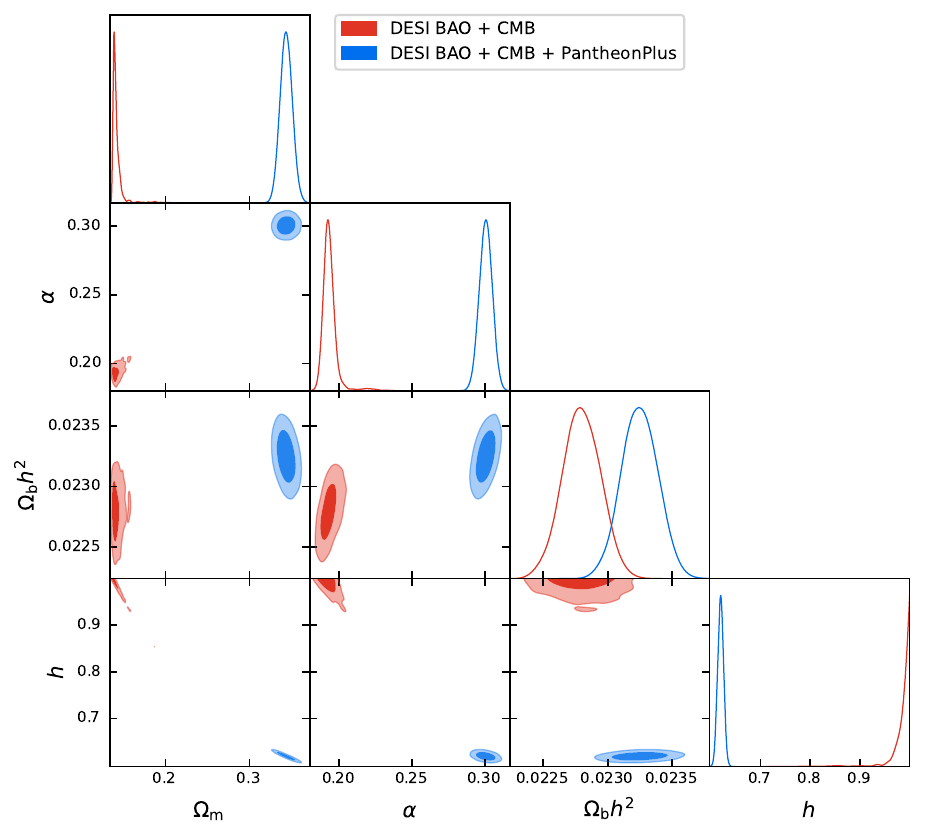}
    \caption{One-dimensional posterior distributions and two-dimensional marginalized contours at 1$\sigma$ and 2$\sigma$ levels for RDE model}
    \label{fig.RDE}
\end{figure}

Fig. \ref{fig.RDE} presents the one-dimensional posterior distributions and two-dimensional marginalized contours of combined observational constraints for the RDE model. Significant discrepancies are observed in the parameter constraints between two datasets. In particular, the red contours of DESI BAO+CMB dataset exceed the specified prior ranges, indicating a substantial mismatch between the model predictions and the observational data. Based on the $\chi^2$ statistic and Bayesian evidence results, which will be discussed in the next chapter, it is found that the RDE model is ruled out by observational data.

The poor performance of the RDE model, which is consistent with previous studies \cite{Xu:2016grp, Wen:2017aaa}, indicates that the Ricci scalar curvature is not a viable choice of IR cutoff. It is worth noting that the RDE model also exhibits the phenomenon of parameter discrepancy. However, in this case, this phenomenon arises primarily from the poor performance of the cosmological fits within the model. 

\subsection{GRDE model}
The third HDE model considered in this paper is the GRDE model, which uses the Granda-Oliver cutoff as the IR cutoff.

\begin{figure}[ht]
    \centering
    \includegraphics[width=0.8\textwidth]{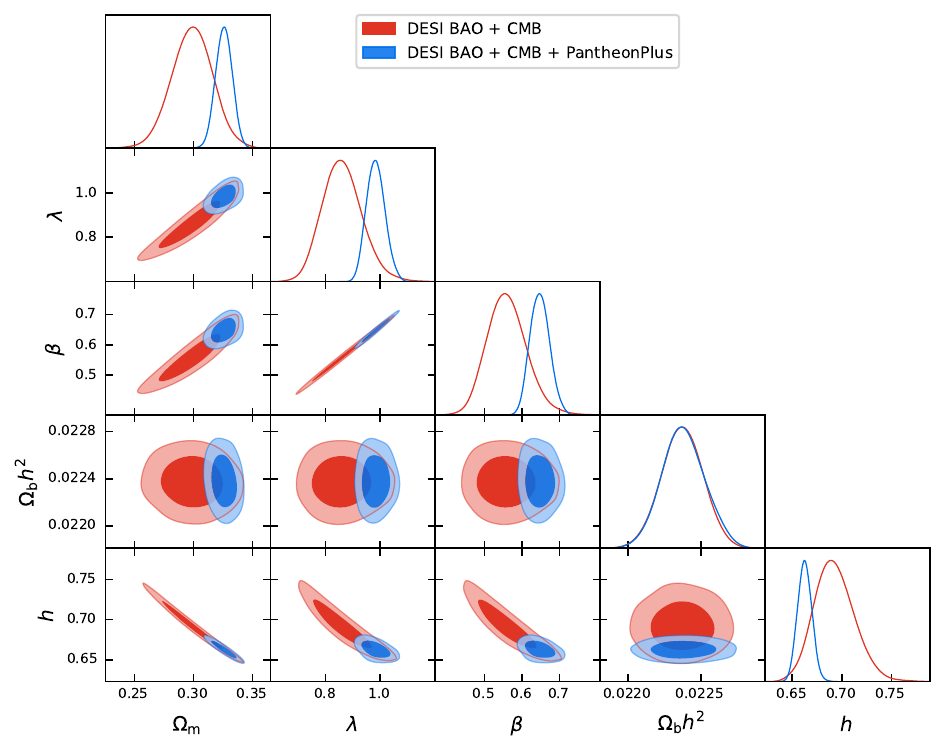}
    \caption{One-dimensional posterior distributions and two-dimensional marginalized contours at 1$\sigma$ and 2$\sigma$ levels for GRDE model}
    \label{fig.GRDE}
\end{figure}

Fig. \ref{fig.GRDE} presents the one-dimensional posterior distributions and two-dimensional marginalized contours of combined observational constraints for the GRDE model. The red and blue contours are generally compatible, with minor shifts observed in parameter distributions when PantheonPlus data is added. Specifically, the addition of SN data slightly tightens constraints on parameters such as $\Omega_m$, $\lambda$, $\beta$, and $h$, as reflected in the narrower blue contours. A noticeable correlation between $\lambda$ and $\beta$ persists in the BAO+CMB+SN dataset. 

The best-fit values derived from the sample likelihood constraints are listed in Table \ref{tab:bfp}. For the DESI BAO+CMB dataset, the best-fit values with $1\sigma$ CL are $\Omega_m=0.2991^{+0.0381}_{-0.0436}$, $h=0.6915^{+0.0563}_{-0.0416}$, $\lambda=0.8539^{+0.1820}_{-0.1466}$, and $\beta=0.5543^{+0.1299}_{-0.1034}$, yielding an EoS of $w=-1.16^{+0.22}_{-0.11}$. In the DESI BAO+CMB+PantheonPlus dataset, we have $\Omega_m=0.3264^{+0.0159}_{-0.0170}$, $h=0.6618^{+0.0181}_{-0.0154}$, $\lambda=0.9907^{+0.0757}_{-0.0775}$, and $\beta=0.6518^{+0.0565}_{-0.0612}$, resulting in $w=-1.00^{+0.08}_{-0.06}$, which brings the EoS of the GRDE model closer to that of $\Lambda$CDM. Based on the parameter constrain results given by the two datasets, the discrepancies for $\Omega_m$, $\lambda$, $\beta$, and $h$ are $0.61 \sigma$, $0.75 \sigma$, $0.74 \sigma$, and $0.57 \sigma$, respectively.

The BAO+CMB datasets give a EoS of GRDE $w<-1$, corresponding to phantom-like characteristics. 
On the contrary, the inclusion of PantheonPlus SN data leads to a EoS of GRDE $w=-1$.
It shows that the PantheonPlus dataset does not favor a dynamical DE.

It is worth noting that, for the GRDE model, the phenomenon of parameter discrepancy is greatly relieved. Compared with the RDE model, the GRDE model introduces one extra parameter, which substantially improves the performance of cosmological fits. This implies that when the Hubble scale or its related combination as the IR cutoff, this parameter discrepancy would not appear. (This conclusion does not hold true for the RDE model, which has already been ruled out by observations.)

\subsection{IHDE1 model}
The fourth HDE model considered in this paper is the IHDE1 model, which incorporates an interaction term and adopts the Hubble horizon as the IR cutoff.

\begin{figure}[ht]
    \centering
    \includegraphics[width=0.8\textwidth]{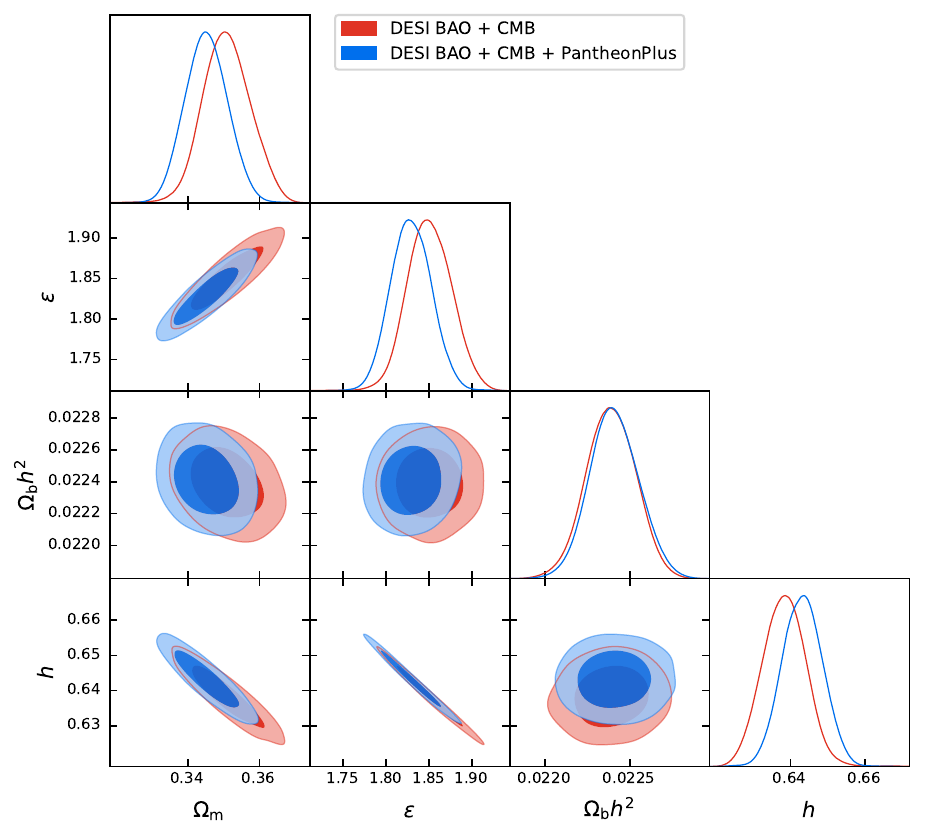}
    \caption{ One-dimensional posterior distributions and two-dimensional marginalized contours at 1$\sigma$ and 2$\sigma$ levels for IHDE1 model}
    \label{fig.IHDE1}
\end{figure}

Fig. \ref{fig.IHDE1} presents the one-dimensional posterior distributions and two-dimensional marginalized contours of combined observational constraints for the IHDE1 model. The red contours represents constraints from the DESI BAO + CMB, while the blue contours represents constraints from the DESI BAO + CMB + PantheonPlus. The constraints are consistent across both datasets, with no significant parameter discrepancies caused by the inclusion of SN data. 

The best-fit values derived from the sample likelihood constraints are listed in Table \ref{tab:bfp}. For the DESI BAO+CMB dataset, the best-fit values with $1\sigma$ CL are $\Omega_m=0.3497^{+0.0156}_{-0.0128}$, $\varepsilon=1.8465^{+0.0618}_{-0.0533}$, and $h=0.6392^{+0.0122}_{-0.0131}$. In the DESI BAO+CMB+SN dataset, we have $\Omega_m=0.3451^{+0.0128}_{-0.0123}$, $\varepsilon=1.8282^{+0.0530}_{-0.0473}$, and $h=0.6433^{+0.0114}_{-0.0117}$. 
Based on the $\chi^2$ statistic and Bayesian evidence results, which will be discussed in the next chapter, it is found that the IHDE1 model is not favored by observational data.

It is clear that, for the IHDE1 model, which chooses the Hubble horizon as the IR cutoff, the phenomenon of parameter discrepancy does not appear. This is very similar to the case of GRDE model.

\subsection{IHDE2 mode}
The fifth HDE model considered in this paper is the IHDE2 model, which incorporates an interaction term and adopts the future event horizon as the IR cutoff.

\begin{figure}[ht]
    \centering
    \includegraphics[width=0.8\textwidth]{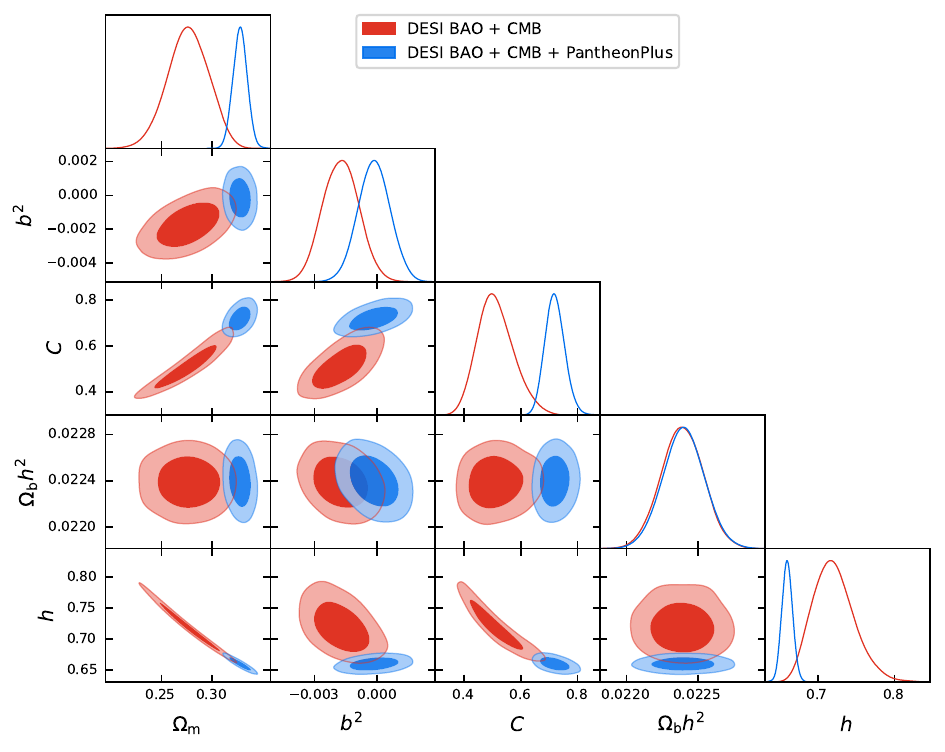}
    \caption{ One-dimensional posterior distributions and two-dimensional marginalized contours at 1$\sigma$ and 2$\sigma$ levels for IHDE2 model}
    \label{fig.IHDE2}
\end{figure}

Fig. \ref{fig.IHDE2} presents the one-dimensional posterior distributions and two-dimensional marginalized contours of combined observational constraints for the IHDE2 model. The overall constraint patterns show similarities to those observed in the OHDE. While the BAO+CMB (red) data provides a broad constraint, the inclusion of PantheonPlus data (blue) generally tightens the parameter constraints, particularly for $\Omega_m$, $b^2$, $C$, and $h$. The blue contours favor higher values of $\Omega_m$, $b^2$, $C$, and lower value of $h$. 

The best-fit values derived from the sample likelihood constraints are listed in Table \ref{tab:bfp}. For the DESI BAO+CMB dataset, the best-fit values with $1\sigma$ CL are $\Omega_m=0.2710^{+0.0487}_{-0.0438}$, $h=0.7263^{+0.0686}_{-0.0587}$, $b^2=-0.0018^{+0.0020}_{-0.0019}$, and $C=0.4913^{+0.1845}_{-0.1124}$, which yields the EoS of $w=-1.48^{+0.34}_{-0.39}$. In the DESI BAO+CMB+PantheonPlus dataset, we have $\Omega_m=0.3279^{+0.0167}_{-0.0165}$, $h=0.6596^{+0.0174}_{-0.0157}$, $b^2=-0.0002^{+0.0017}_{-0.0017}$, and $C=0.7171^{+0.0867}_{-0.0704}$, resulting in $w=-1.09^{+0.09}_{-0.09}$. Based on the parameter constrain results given by the two datasets, the discrepancies for $\Omega_m$, $b^2$, $C$, and $h$ are $1.15\sigma$, $0.61\sigma$, $1.34\sigma$, and $1.01\sigma$, respectively.

Both two datasets give a EoS $w<-1$, indicating that the IHDE2 model also has the phantom-like characteristics. 
The best-fit coupling $b^2$ is very close to zero but remains negative, suggesting a weak interaction with potential energy transfer from the matter to DE.
Similar to the OHDE model, the EoS of the IHDE2 model deviates significantly from the $\Lambda$CDM case ($w=-1$) when only the BAO+CMB data are taken into account. However, the inclusion of the SN data leads to the EoS of IHDE2 much closer to the cosmological constant. This implies that the PantheonPlus dataset does not favor a dynamical DE.

Once again, a significant parameter discrepancy is observed for the IHDE2 model. This phenomenon can be attributed to the choice of the future event horizon as the IR cutoff. This is very similar to the case of OHDE model.

\subsection{THDE model}
The sixth HDE model considered in this paper is the THDE model, which considered the Tsallis entropy and adopts the Hubble horizon as the IR cutoff.

\begin{figure}[ht]
    \centering
    \includegraphics[width=0.8\textwidth]{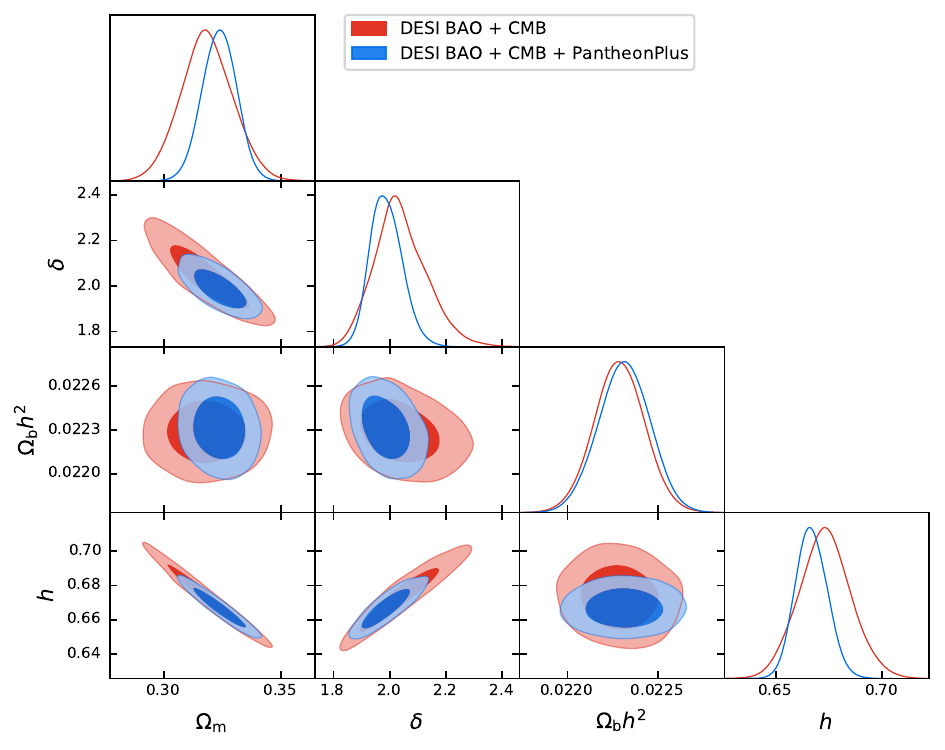}
    \caption{ One-dimensional posterior distributions and two-dimensional marginalized contours at 1$\sigma$ and 2$\sigma$ levels for THDE model}
    \label{fig.THDE}
\end{figure}

For our analysis, we fixed $B=3$ in $M_P^2$ units to perform constraints. Fig. \ref{fig.THDE} presents the one-dimensional posterior distributions and two-dimensional marginalized contours of combined observational constraints for the THDE model. The parameter constraints are consistent across both datasets, with no deviations caused by the inclusion of SN data. The addition of SN data slightly tightens constraints on parameters such as $\Omega_m$, $\delta$, and $h$, as reflected in the narrower blue contours.

The best-fit values derived from the sample likelihood constraints are listed in Table \ref{tab:bfp}. For the DESI BAO+CMB dataset, the best-fit values with $1\sigma$ CL are $\Omega_m=0.3200^{+0.0237}_{-0.0245}$, $h=0.6710^{+0.0281}_{-0.0245}$, and $\delta=2.0211^{+0.2430}_{-0.1732}$, which yields the EoS of $w=-1.00^{+0.06}_{-0.05}$. In the DESI BAO+CMB+PantheonPlus dataset, we have $\Omega_m=0.3233^{+0.0164}_{-0.0138}$, $h=0.6666^{+0.0153}_{-0.0151}$, and $\delta=1.9818^{+0.1261}_{-0.1047}$, resulting in $w=-0.99^{+0.03}_{-0.03}$. The evolution behavior of THDE is very close to $\Lambda$CDM.

Since the THDE model chooses the Hubble horizon as the IR cutoff, the phenomenon of parameter discrepancy does not appear. This is very similar to the cases of the GRDE and IHDE1 models.

\subsection{BHDE model}
The seventh HDE model considered in this paper is the BHDE model, which considered the Barrow entropy and adopts the future event horizon as the IR cutoff.

For consistency with the THDE, we fixed $B=3$ in $M_P^2$ units to perform constraints. When $\Delta =0$, this model reduces to the OHDE scenario.

\begin{figure}[ht]
    \centering
    \includegraphics[width=0.8\textwidth]{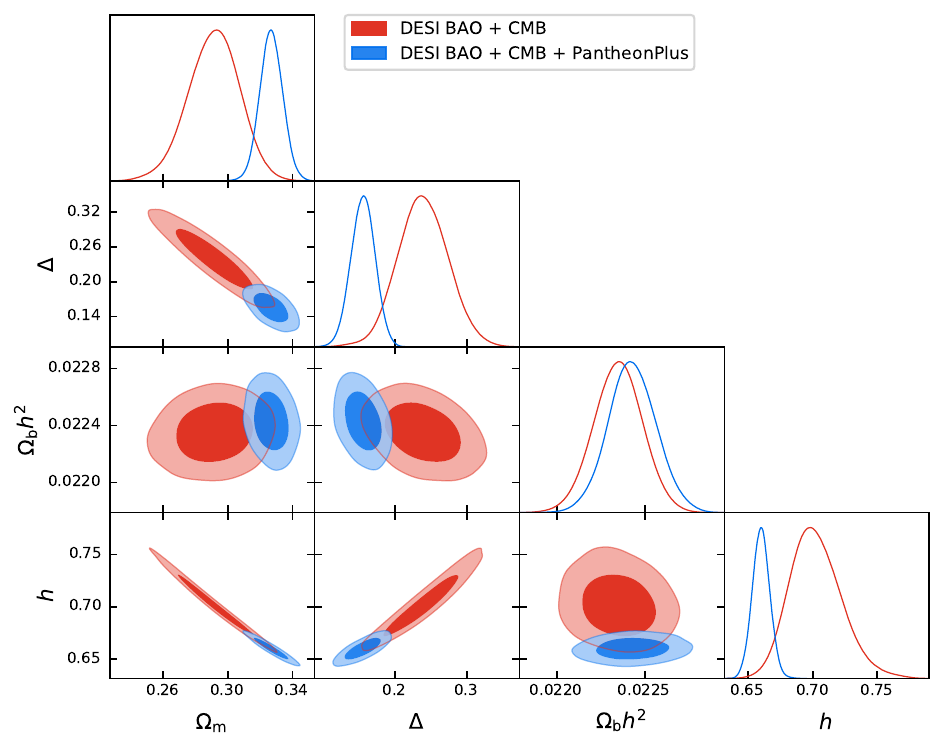}
    \caption{ One-dimensional posterior distributions and two-dimensional marginalized contours at 1$\sigma$ and 2$\sigma$ levels for BHDE model}
    \label{fig.BHDE}
\end{figure}

Fig. \ref{fig.BHDE} presents the one-dimensional posterior distributions and two-dimensional marginalized contours of combined observational constraints for the BHDE model. The overall constraint patterns show similarities to those observed in the OHDE and IHDE2 models, both of which use the future event horizon as a cutoff. While the DESI BAO+CMB (red) data provides a broad constraint, the inclusion of PantheonPlus data (blue) generally tightens the parameter constraints, particularly for $\Omega_m$, $\Delta$, and $h$. The blue contours favor higher value of $\Omega_m$, and lower values of $\Delta$, and $h$. In addition, there is a noticeable discrepancy between the results for the blue contours and those for the red contours. 

The best-fit values derived from the sample likelihood constraints are listed in Table \ref{tab:bfp}. For the DESI BAO+CMB dataset, the best-fit values with $1\sigma$ CL are $\Omega_m=0.2921^{+0.0321}_{-0.0338}$, $h=0.7000^{+0.0459}_{-0.0361}$, and $\Delta=0.2349^{+0.0740}_{-0.0724}$, which yields an EoS of $w=-1.26^{+0.11}_{-0.18}$ at $z=0$. In the DESI BAO+CMB+SN dataset, we have $\Omega_m=0.3261^{+0.0157}_{-0.0137}$, $h=0.6606^{+0.0130}_{-0.0147}$, and $\Delta=0.1581^{+0.0298}_{-0.0379}$, resulting in $w=-1.09^{+0.04}_{-0.04}$ at $z=0$. Based on the parameter constrain results given by the two datasets, the discrepancies for $\Omega_m$, $\Delta$, and $h$ are $0.94\sigma$, $0.95\sigma$, and $0.91\sigma$, respectively.

Both datasets yield a EoS of BHDE $w<-1$, indicating that BHDE also have phantom-like characteristics. Similar to the OHDE models, the EoS of the BHDE model deviates significantly from the $\Lambda$CDM case ($w=-1$) when only the BAO+CMB data are considered. However, the inclusion of the SN data leads to the EoS of BHDE much closer to the cosmological constant. 
Again, this implies that under the framework of HDE with future event horizon as the IR cutoff, the PantheonPlus dataset does not favor a dynamical DE.

Since the BHDE model chooses the future event horizon as the IR cutoff, it also exhibits a significant parameter discrepancy. Based on the constraint results of the OHDE, IHDE2, and BHDE models, it can be concluded that the occurrence of parameter discrepancy is driven by the choice of the future event horizon as the IR cutoff.

\section{Conclusion}
\label{sec:conclusion}
The results of $\chi^2$ statistic and Bayesian evidence for the two datasets are summarized in Table \ref{tab:BAOCMB} and Table \ref{tab:BAOCMBSN}, respectively. 

\begin{table}
    \centering
    \begin{tabular}{lccc}
    \hline \hline
    \multirow{2}{*}{} & \multicolumn{3}{c}{\textbf{DESI BAO+CMB}} \\ \cline{2-4}
    \textbf{Model} & \textbf{$\chi^2_{\text{min}}$} & \textbf{$\ln \mathcal{B}_{12}$} & \textbf{Rank} \\ \hline
    \textbf{$\Lambda$CDM} & 21.22  & 0      & 1 \\
    \textbf{OHDE}         & 21.61  & -2.18  & 2 \\
    \textbf{THDE}         & 21.20  & -2.39  & 2 \\
    \textbf{BHDE}         & 20.86  & -3.34  & 2 \\
    \textbf{GRDE}         & 16.87  & -5.80  & 2 \\
    \textbf{IHDE2}        & 22.55  & -9.55  & 3 \\
    \textbf{IHDE1}        & 40.45  & -13.74 & 3 \\
    \textbf{RDE}          & 183.70 & -89.61 & 4 \\
    \hline \hline
    \end{tabular}
    \caption{The $\chi^2$ statistic values and relative Bayesian evidence for each model in BAO+CMB dataset. The models follow the relative Bayesian evidence in descending order.}
    \label{tab:BAOCMB}
\end{table}

\begin{table}[h]
    \centering
    \begin{tabular}{lccc}
    \hline \hline
    \multirow{2}{*}{} & \multicolumn{3}{c}{\textbf{DESI BAO+CMB+PantheonPlus}} \\ \cline{2-4}
    \textbf{Model} & \textbf{$\chi^2_{\text{min}}$} & \textbf{$\ln \mathcal{B}_{12}$} & \textbf{Rank} \\ \hline
    \textbf{$\Lambda$CDM} & 1424.43 & 0       & 1 \\
    \textbf{THDE}         & 1424.28 & -3.29   & 2 \\
    \textbf{BHDE}         & 1431.62 & -7.61   & 2 \\
    \textbf{GRDE}         & 1422.87 & -8.35   & 2 \\
    \textbf{OHDE}         & 1436.45 & -9.78   & 2 \\
    \textbf{IHDE2}        & 1435.45 & -15.15  & 3 \\
    \textbf{IHDE1}        & 1452.89 & -18.81  & 3 \\
    \textbf{RDE}          & 1997.47 & -291.46 & 4 \\
    \hline \hline
    \end{tabular}
    \caption{The $\chi^2$ statistic values and relative Bayesian evidence for each model in BAO+CMB+PantheonPlus dataset. The models follow the relative Bayesian evidence in descending order.}
    \label{tab:BAOCMBSN}
\end{table}

Based on Bayesian evidence, we make a rank for all the DE models considered in this paper. For the case of the BAO+CMB dataset, the $\Lambda$CDM is in the first rank since it yields the best Bayesian evidence among all models. The OHDE, THDE, BHDE, and GRDE are classified in the second rank, with their Bayesian evidence greater than $-6$. The IHDE2 and IHDE1 are placed in the third rank.
Finally, the RDE, with the highest $\chi^2$ values and the lowest Bayesian evidence, is positioned in the fourth rank.

Different from the results of the BAO+CMB dataset, minor changes appear for the case of the BAO+CMB+SN dataset. 
The $\Lambda$CDM remains in the first rank. However, the order of the models in the second rank becomes: THDE, BHDE, GRDE and OHDE, with their Bayesian evidence greater than $-10$. Due to the low Bayesian evidence, the IHDE2 and IHDE1 are still classified into the third rank. 
The RDE, which is ruled out by observation, is placed in the fourth rank.

Moreover, we draw the following key conclusions:

1. Based on Bayesian evidence, the $\Lambda$CDM remains the most competitive model, while the RDE model is ruled out by observational data.

2. HDE models with dark sector interaction perform the worst across all categories, indicating that the interaction term is not favored within the HDE framework. 

3. The remaining three categories show relatively comparable performance. Specifically, the OHDE model has better performance in the BAO+CMB dataset, while the HDE models with modified black hole entropy outperforms other three categories in the BAO+CMB+SN dataset.

4. The key factor responsible for the parameter discrepancy is the choice of the characteristic length $L$. HDE models with the future event horizon as the IR cutoff exhibit significant discrepancies in parameter constraints between the DESI BAO+CMB and DESI BAO+CMB+PantheonPlus datasets. The BAO+CMB dataset favors a phantom-like DE, whereas the inclusion of PantheonPlus data brings the EoS much closer to the cosmological constant. This indicates that, in the framework of HDE with the future event horizon, the PantheonPlus dataset does not favor a dynamical DE.

\section{Summary}
\label{sec:Summary}

In this paper, we perform a comprehensive numerical study on all four categories of HDE. Seven representative HDE models across four categories are selected, including OHDE, RDE, GRDE, IHDE1, IHDE2, THDE, and BHDE. Among these models, the GRDE and IHDE2 introduce two additional free parameters compared to the $\Lambda$CDM, while the remaining models introduced only one additional parameter. For comparison, we adopt the $\Lambda$CDM model as the fiducial model. The observational data include DESI BAO 2024 measurements, Planck 2018 CMB distance priors, and the PantheonPlus compilation of SN. These data are divided into two sets: DESI BAO+CMB, and DESI BAO+CMB+PantheonPlus. We apply $\chi^2$ statistic to evaluate the models' compatibility with current observational data, and then compare their relative performances with Bayesian evidence.

Based on Bayesian evidence, we find that these seven HDE models can be classified into four rank. 
The $\Lambda$CDM model is in the first rank, the OHDE, THDE, BHDE, and GRDE models are in the second rank, 
the IHDE2 and IHDE1 models are in the third rank, and the RDE is in the forth rank.

In addition, the inclusion of SN data introduces discrepancies in the performance of different HDE models. 
For the HDE models that adopt the Hubble scale as a IR cutoff, their EoS are very close to the cosmological constant, 
and the inclusion of PantheonPlus dataset does not affect the results. 
For HDE models that adopt the extended Hubble scale as a IR cutoff, their EoS exhibit phantom-like characteristics, 
and the inclusion of PantheonPlus dataset leads to slight discrepancies in the parameter constraints, thus bringing the EoS closer to $-1$.
Finally, for the HDE models that adopt the future event horizon as a cutoff, their EoS display obvious phantom-like characteristics, 
and the inclusion of PantheonPlus dataset significantly alters parameters constraints, also bringing the EoS closer to $-1$.

Our studies imply that, under the framework of HDE with future event horizon, the PantheonPlus dataset does not favor a dynamical DE.
This indicates the importance of further investigating PantheonPlus SN data. 
It is interesting to explore the details of PantheonPlus SN samples, 
and then identify which redshift range plays a more important role in favoring the cosmological constant. 
Furthermore, instead of focusing on a specific class of DE models, one can also adopt model-independent methods to analyze observational data. 
These topics are worth further explorations in future works.

\acknowledgments

We want to thank the anonymous referee, whose valuable suggestions greatly improves the quality of our work. J.X. Li wants to thank Yi Zheng for helpful discussions about DESI BAO data. S.Wang is supported by a grant of Guangdong Provincial Department of Science and Technology under No. 2020A1414040009.

\newpage
% Bibliography
\bibliographystyle{JHEP}
\bibliography{paper.bib}

\end{document}